\documentclass[aps,superscriptaddress,twocolumn,a4paper,10pt]{revtex4}
\usepackage[colorlinks=true, pdfstartview=FitV, linkcolor=blue, citecolor=red, urlcolor=black]{hyperref}
\usepackage{graphicx}
\usepackage[all]{xy}
\usepackage{amsmath}
\usepackage{amssymb}
\usepackage{tensor}
\usepackage{orcidlink}
\usepackage{amsthm}
\usepackage{comment}
\newcommand{\be}{\begin{equation}}
\newcommand{\ee}{\end{equation}}
\newcommand{\ben}{\begin{eqnarray}}
\newcommand{\een}{\end{eqnarray}}
\newcommand{\bes}{\begin{subequations}}
\newcommand{\ees}{\end{subequations}}
\def\bal#1\eal{\begin{align}#1\end{align}}
\newcommand{\ov}{\overline}

\newcommand{\bfi}{\begin{figure}}
\newcommand{\efi}{\end{figure}}
\newcommand{\bc}{\begin{center}}
\newcommand{\ec}{\end{center}}

\newcommand{\vphia}{|\varphi|}
\newcommand{\vphi}{{\varphi}}
\newcommand{\sech}{{\rm sech}}
\newcommand{\csch}{{\rm csch}}

\newcommand{\arcsinh}{{\rm arcsinh}}

\newcommand{\LL}{{\cal L}}

\newcommand{\Sc}{{\cal S}}

\begin{document}

\title{Super long-range vortices}
\author{C.A.G Almeida\,\orcidlink{0000-0001-7649-7138}}
     \email[]{cviniro@dcx.ufpb.br}\affiliation{Departamento de Ci\^encias Exatas, Universidade Federal
da Para\'{\i}ba, 58297-000 Rio Tinto, PB, Brazil}
\author{I. Andrade\,\orcidlink{0000-0002-9790-684X}}
        \email[]{andradesigor0@gmail.com}\affiliation{Departamento de F\'\i sica, Universidade Federal da Para\'\i ba, 58051-970 Jo\~ao Pessoa, PB, Brazil}
\author{M.A. Marques\,\orcidlink{0000-0001-7022-5502}}
        \email[]{marques@cbiotec.ufpb.br}\affiliation{Departamento de Biotecnologia, Universidade Federal da Para\'\i ba, 58051-900 Jo\~ao Pessoa, PB, Brazil}
\author{R. Menezes\,\orcidlink{0000-0002-9586-4308}}
     \email[]{rmenezes@dcx.ufpb.br}\affiliation{Departamento de Ci\^encias Exatas, Universidade Federal
da Para\'{\i}ba, 58297-000 Rio Tinto, PB, Brazil}

\begin{abstract}
In this work, we investigate the presence of vortex configurations with logarithmic tails, which we call super long-range vortices, in Maxwell-Higgs models with gauge field dynamics modified by generalized magnetic permeability in the Lagrangian density. By taking advantage of a first-order formalism, we study which behavior the magnetic permeability must have in order to allow for the presence of the logarithmic tails in the solutions. We also analyze the asymptotic behavior of the magnetic field and energy density. To illustrate our procedure, we present two models; one of them is described by analytical solutions.
\end{abstract} 

\maketitle

\section{Introduction}
Vortices are topological structures that appear in Field Theory under the action of a scalar field coupled to a gauge field through a local $U(1)$ symmetry in $(2,1)$ flat spacetime dimensions, whose standard relativistic model was proposed by Nielsen and Olesen \cite{NO}. A first-order formalism compatible with the equations of motion was obtained in Ref.~\cite{bogo}. Notwithstanding that, the non-linear character of the equations do not allow for the presence of analytical solutions, so one must use numerical procedures to obtain them \cite{schaposnik}.

The minimum energy vortex configurations that arise in the model originally proposed in \cite{NO} require that the potential must be the so-called $\vphia^4$; they engender magnetic field and energy density with disk-like shape. To obtain vortices with different features, one must consider non-canonical models with modifications on the dynamical term of the scalar \cite{vortexM1,vortexM2} and/or of the gauge field \cite{mu1,mu2,mu3,longrange1,longrange2,longrange3,ring1,ring2,mu4,mu5,mu56,mu6,mu7,mu8}. In the latter case, the scalar field is used to modify the magnetic permeability of the system. These modifications allow for obtaining distinct profiles, such as compact \cite{mu56} and ring-like \cite{ring1,ring2} vortices. Other classes of generalized models, such as $k$-gauge fields and Born-Infeld, were investigated in \cite{vortexY1,vortexY2,vortexY3,vortexY4,vortexY5,vortexY6,vortexY7,vortexY8}. Interestingly, under some conditions, which depend on the specific model studied, one may also obtain first-order equations which satisfy Derrick's rescaling argument \cite{derrick}; see Ref.~\cite{vortexY7}.

The Nielsen-Olesen vortex engenders asymptotic behavior described by mix of power-law and exponential functions. An interesting possibility that appears due to the presence of generalized magnetic permeability is the 
long-range vortices \cite{longrange3}, whose falloff is slower than the standard ones. It is worth highlighting that the study of kinks, which are other topological structures, with long-range profile was considered by several works \cite{longkink1,longkink2,longkink3,longkink4,longkink5,longkink6} due to their highly interactive behavior. Since long-range structures engender tails which extend farther than the short-range ones, their interactions and collisions are of current interest \cite{colisao1,colisao2,colisao3,colisao4,colisao5,colisao6,colisao7,colisao8,colisao9}.

In the recent papers \cite{superlongkink1,superlongkink2}, the authors presented scalar field models that support kinks with logarithmic asymptotic behavior, so they were called super long-range kinks. There, these structures arise due to an exotic feature of the potential: all orders of its derivatives vanish. In this manuscript, we investigate vortices with logarithmic tails in Maxwell-Higgs models with generalized magnetic permeability. In Sec.~\ref{secmodel}, we present the model and the general equations of interest for the study. Using this general model, we discuss which conditions the magnetic permeability must obey in order to admit solutions with logarithmic tails, i.e., super long-range vortices. In Sec.~\ref{secfirst}, we present a model inspired by recent results in kinks; in this case, the first-order equations cannot be solved analytically, requiring a numerical approach to obtain the solutions. In Sec.~\ref{secsecond}, we generalize the logarithmic decay, considering a model which supports analytical solutions containing a parameter that control the intensity of the super long-range behavior. In Sec.~\ref{secoutlook}, we present our final remarks and discuss perspectives for future research.

\section{Maxwell-Higgs model}\label{secmodel}
We consider the action of a complex scalar field $\vphi$ coupled to a gauge field $A_\alpha$ via $U(1)$ symmetry in $(2,1)$ flat spacetime dimensions with metric $\eta_{\alpha\beta}=\text{diag}(+,-,-)$. The action has the form $\Sc=\int d^3x\,\LL$ and the Lagrangian density of our interest engenders a generalized magnetic permeability $\mu(|\vphi|)$ driven by the scalar field, as
\be\label{modelM}
\LL = -\frac{1}{4\mu(\vphia)}F_{\alpha\beta}F^{\alpha\beta} +\ov{D_\alpha\vphi}D^\alpha\vphi - V(\vphia),
\ee
where $F_{\alpha\beta} = \partial_\alpha A_\beta - \partial_\beta A_\alpha$ is the electromagnetic strength tensor and $D_\alpha = \partial_\alpha + iA_\alpha$ denotes the covariant derivative. The potential is denoted by $V(\vphia)$, which is supposed to support a set of minima at $\vphia = v$, where $v$ is a non-null parameter associated to symmetry breaking.

The non-trivial magnetic permeability ($\mu\neq1$) allows for the presence of distinct features, such as ring-like shaped or long-range vortices; see, for instance, Refs.~\cite{mu1,mu2,mu3,longrange1,longrange2,longrange3,ring1,ring2,mu4,mu5,mu56,mu6,mu7,mu8}. In this paper, we unveil vortices whose tails fall off logarithmically. The general formalism was previously obtained in the aforementioned references, so we go straight to the point in which one considers static fields and $A_0=0$, with the following ansatz
\be\label{ansatz}
\vphi = g(r)e^{in\theta} \quad\text{and}\quad \vec{A} = \frac{\hat{\theta}}{r}\left(n -a(r)\right),
\ee
where $(r,\theta)$ represents the polar coordinates and $n = \pm1, \pm2, \pm3,\dots$ is the vorticity. For this choice, the fields exhibit circular symmetry and are described by $g(r)$ and $a(r)$, with boundary conditions
\be\label{bound}
\begin{aligned}
g(0) &=0, \quad\quad a(0) =n,\\
\lim_{r\to\infty}g(r) &=1, \quad\quad \lim_{r\to\infty}a(r) =0.
\end{aligned}
\ee
In this case, the magnetic field $B = -F^{12} = -\varepsilon^{ij} \partial_i A_j$ reads
\be\label{bfield}
B = -\frac{a'}{r}.
\ee
By integrating it, one obtain the quantized flux $\Phi = 2\pi n$, which can be either positive or negative. The prime denotes a derivative with respect to the coordinate $r$.

By varying the action associated to \eqref{modelM} with respect to the scalar and gauge fields, we get the equations of motion
\bes\label{eom}\bal\label{eomPaz}
&\frac{1}{r}\left(rg^\prime\right)^\prime -\frac{a^2g}{r^2} +\frac{\mu_g{a^\prime}^2}{4r^2\mu^2} -\frac12V_g = 0,\\
\label{eomAaz}
&\left(\frac{a^\prime}{r\mu}\right)^\prime -\frac{2ag^2}{r} = 0,
\eal
\ees
in which we have used the ansatz \eqref{ansatz}.

The energy density is obtained as usual, by considering $\rho=-\LL$, which becomes
\be\label{rhoaz}
\rho = \frac{{a^\prime}^2}{2r^2\mu} +{g^\prime}^2 +\frac{a^2g^2}{r^2} +V.
\ee
The energy can be calculated by integrating the above expression, $E=\int_{\mathbb{R}^2} d^2x \rho = 2\pi\int^\infty_0 r\,dr\,\rho$. The boundary conditions \eqref{bound} ensure that the energy of the solutions is finite. However, obtaining solutions is not an easy task, as one must solve the equations of motion \eqref{eom}, which are of second order with the presence of nonlinearity and couplings between $a(r)$ and $g(r)$. In Refs.~\cite{mu1,mu2,mu3}, one finds a first-order formalism based on the minimization of the energy which is compatible with the equations of motion. It requires that the potential is related to the magnetic permeability via
\be\label{pot}
V(\vphia) = \frac12\mu(\vphia)\left(1-\vphia^2\right)^2.
\ee
In this situation, one can show that the first-order equations
\be\label{foaz}
g^\prime = \pm\frac{ag}{r} \quad\text{and}\quad -\frac{a^\prime}{r} = \pm\mu(g)\left(1-g^2\right)
\ee
can be used to obtain vortex solutions with the minimum energy of the system, $E =  2\pi |n|$, complying with the boundary conditions \eqref{bound}. It is worth commenting that, solutions obeying the above first-order equations engender null stress; see Ref.~\cite{vortexY7}. Interestingly, the behavior of $g(r)$ near the origin is always $g(r)\propto r^n$, whilst $a(r)$ depends on the form of $\mu(g)$. The equations with upper and lower signs are related with $a \to -a$. For simplicity, we consider only the ones with upper sign and $n = 1$.

Since our interest is to obtain vortex solutions with distinct tails, let us briefly review the behavior in the standard case, $\mu=1$ \cite{NO,bogo,schaposnik}. In this case, the potential is the well-known $\vphia^4$, given by $V(\vphia) = \frac12\left(1-\vphia^2\right)^2$. The asymptotic behavior of the functions $g(r)$ and $a(r)$ is
\be\label{asystandard}
1-g(r) \approx \frac{\lambda e^{-\sqrt{2}\,r}}{\sqrt{r}} \quad\text{and}\quad
a(r) \approx\lambda\sqrt{2r}e^{-\sqrt{2}\,r},
\ee
in which $\lambda$ is a constant of integration. Notice that there is no freedom to modify the above behavior in the standard model. In order to do so, one must consider generalized models. In this direction, we include the magnetic permeability $\mu(|\vphi|)$, in the form
\be\label{mulong}
\mu(|\vphi|) \approx \frac{1}{2\kappa^2}\left(1-|\vphi|^2\right)^{2/\ell},
\ee
with $\ell>0$, around points where the field approaches the set of minima, $\vphia\approx1$. The models investigated in Ref.~\cite{longrange3} fall within the above expression, which leads to long-range vortices, whose solutions engender power-law tails, such that the asymptotic behavior is given by
\be\label{asylong}
g(r) \approx 1-\frac{\ell^\ell\kappa^\ell}{2}\,\frac{1}{r^{\ell}} \quad\text{and}\quad a(r) \approx \frac{\ell^{\ell+1}\kappa^\ell}{2}\,\frac{1}{r^{\ell}}.
\ee
This behavior has also appeared in Refs.~\cite{longrange1,longrange2,longrange3}. As $\ell$ gets larger and larger, the vortex extends farther and farther.

\subsection{Super long-range tails}
Our goal is to obtain a new class of models that supports vortices whose solutions fall off even slower than the power-law ones. In Ref.~\cite{superlongkink2}, the authors have introduced the scalar field model $\LL = \frac12\partial_\mu\phi\partial^\mu\phi - \frac12\cos^4(\phi)\sech^2(a\tan(\phi))$ in $(1,1)$ dimensions that admits solutions with asymptotic behavior $\phi \approx \pi/2 - a/\ln(2ax)$, which were called super long-range kinks. The interaction between these structures are stronger than the short- and long-range ones.

To seek vortices with super long-range asymptotic behavior, we analyze the first-order equations \eqref{foaz} to verify how the magnetic permeability $\mu(\vphia)$ in \eqref{modelM} must behave. The solutions formed by the pair $g(r)$ and $a(r)$ engender logarithmic tails if the permeability has the form
\be\label{musuper}
\mu(|\vphi|) \approx \alpha^2\left(1-|\vphi|^2\right)^{2/\gamma}\,e^{-\beta\sqrt{2\gamma(\gamma+1)}/|1-|\vphi|^2|^{1/\gamma}}
\ee
around the points in which $\vphia\approx1$, where $\alpha$ is positive, and both $\beta$ and $\gamma$ are non-negative real parameters. In this situation, for $\beta$ and $\gamma$ positive, we have the asymptotic behavior
\be\label{asysuper}
g(r)\approx 1 -\frac{\vartheta}{\ln^{\gamma}(\alpha\beta r)}\quad\text{and}\quad
a(r)\approx \frac{\gamma\vartheta}{\ln^{\gamma+1}(\alpha\beta r)},
\ee
in which $\vartheta=(2^{-\frac{\gamma+2}{\gamma}}\beta^2\gamma(\gamma+1))^{\gamma/2}$. Notice that the functions $g(r)$ and $a(r)$, which drive the fields accordingly to the ansatz \eqref{ansatz}, have both a logarithmic decay whose intensity is controlled by the positive real parameter $\gamma$. Due to this feature, we call structures described by solutions with tails \eqref{asysuper} \emph{super long-range vortices}. We remark that $\beta=0$ recovers the magnetic permeability \eqref{mulong} with $\alpha^2=1/(2\kappa^2)$ and $\gamma=\ell$, so the expression \eqref{asysuper} is not valid anymore. Instead, this case leads to the power law tails shown in Eq.~\eqref{asylong}. Moreover, the situation with $\beta=0$ and $\gamma\to\infty$ recovers the standard model, as $\mu(\vphia)\approx\alpha^2$ becomes constant that can be absorbed by redefining fields and coordinates; in this situation, the asymptotic behavior is \eqref{asystandard}. Therefore, if the Lagrangian density \eqref{modelM} supports a magnetic permeability that behaves as \eqref{musuper} at $\vphia\approx1$, one obtains a model which leads to super long-range vortices and also encompasses the standard and long-range profiles. We remark that, even though the solution engenders logarithmic asymptotics \eqref{asysuper}, the magnetic field \eqref{bfield} and the energy density \eqref{rhoaz} fall off as
\bal \label{basysuper}
B(r) &\approx \frac{\gamma(\gamma+1)\vartheta}{r^2\ln^{\gamma+2}(\alpha\beta r)},\\ \label{rhoasysuper}
\rho(r) &\approx \frac{2\gamma(2\gamma+1)\vartheta^2}{r^2\ln^{2(\gamma+1)}(\alpha\beta r)},
\eal
which mix power-law and logarithmic functions.

It is worth commenting that, even though the behavior of $\mu(\vphia)$ around $\vphia\approx1$ must be of the form \eqref{musuper} to allow for the presence of super long-range vortices, finding its general form is not straightforward. Next, we unveil two models with generalized magnetic permeability that support super long-range vortices.

\section{First model}\label{secfirst}
To obtain vortex solutions with asymptotic behavior \eqref{asysuper}, we get inspiration from \cite{superlongkink2}, where the authors considered non-analytic potentials. There, the presence of a hyperbolic secant in the potential makes its derivatives vanish in all orders. With this motivation, we investigate the model \eqref{modelM} with the magnetic permeability given by
\be\label{munum1}
\mu(\vphia) = \left(1-\vphia^2\right)^2\sech^2\left(\frac{b\,\vphia^2}{1-\vphia^2}\right),
\ee
for $\vphia\neq1$, and $\mu(1)=0$. The real parameter $b$ is non negative. We can expand the above expression around $\vphia=1$ to show that it is compatible with \eqref{musuper} for
\be\label{correspondencianum}
\alpha=2,\quad \beta=b \quad\text{and}\quad\gamma=1.
\ee
Since we are interested in minimum-energy stable solutions, we use the first-order formalism, which requires that the potential has the form \eqref{pot}, which reads
\be\label{potnum}
V(\vphia) = \frac12\left(1-\vphia^2\right)^4\sech^2\left(\frac{b\,\vphia^2}{1-\vphia^2}\right)
\ee
for $\vphia\neq1$, and $V(1)=0$. It supports maximum at $\vphia=0$ and a set of minima at $\vphia=1$. For $b\neq0$, all the derivatives of $V(\vphia)$ are null at its minima, i.e., $d^k V(\vphia)/d\vphia^k =0$ at $\vphia=1$, with $k\in\mathbb{N}$. 

The first-order equations \eqref{foaz} can be written as
\be\label{fonum}
g^\prime = \frac{ag}{r} \quad\text{and}\quad -\frac{a^\prime}{r} = \left|1-g^2\right|^3\sech^2\left(\frac{b\,g^2}{1-g^2}\right).
\ee
We were not able to find analytical solutions of the above equations. However, by analyzing them near the points where $g(r)\approx1$ and $a(r)\approx0$, one can show that the asymptotic behavior is
\be\label{solasynum}
g(r) \approx 1 -\frac{b}{2\ln(2br)} \quad\text{and}\quad
a(r) \approx \frac{b}{2\ln^2(2br)},
\ee
for $b\neq0$. Therefore, as expected, the vortex solution associated to \eqref{munum1} engender super long-range tails, falling off logarithmically. The associated magnetic field \eqref{bfield} and the energy density \eqref{rhoaz} behave asymptotically as
\be\label{brhoasynum}
B(r) \approx \frac{b}{r^2\ln^3(2br)},\quad
\rho(r) \approx \frac{3b^2}{2r^2\ln^4(2br)},
\ee
for $b\neq0$, matching with the expressions \eqref{basysuper} and \eqref{rhoasysuper} for the parameters in \eqref{correspondencianum}. We emphasize that the expressions \eqref{solasynum} and \eqref{brhoasynum} are only valid for $b\neq0$. The case $b=0$ is special, as the asymptotic behavior becomes $g(r) \approx 1 -\sqrt{2}/(4r)$ and 
$a(r) \approx \sqrt{2}/(4r)$, so the associated magnetic field $B(r) \approx \sqrt{2}/(4r^3)$ and energy density $\rho(r) \approx 1/(2r^4)$ are of the power-law type. So, the super long-range profile only appears for $b\neq0$. We have checked by numerical integration that the value of the energy is the one expected from the first-order formalism with $n=1$, $E=2\pi$.

\begin{figure}[t!]
\centering
\includegraphics[width=0.5\linewidth]{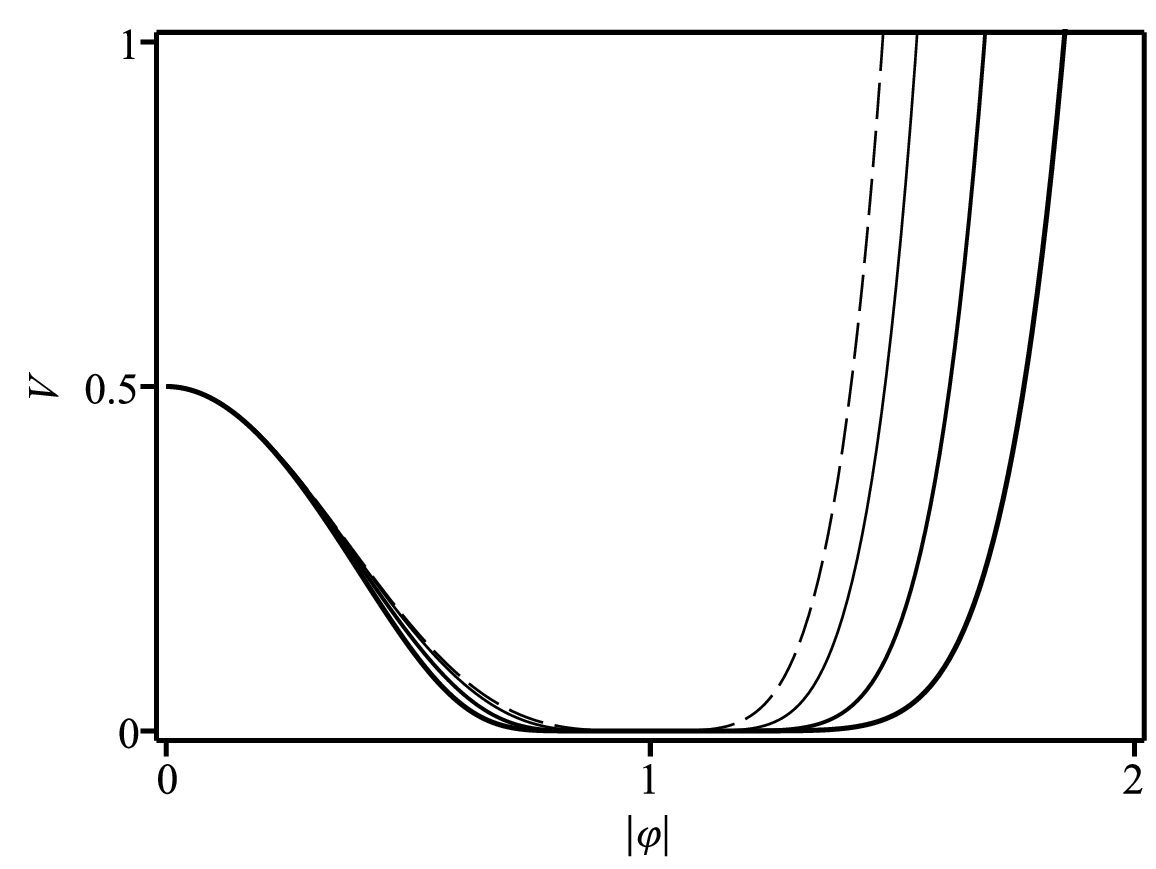}\includegraphics[width=0.5\linewidth]{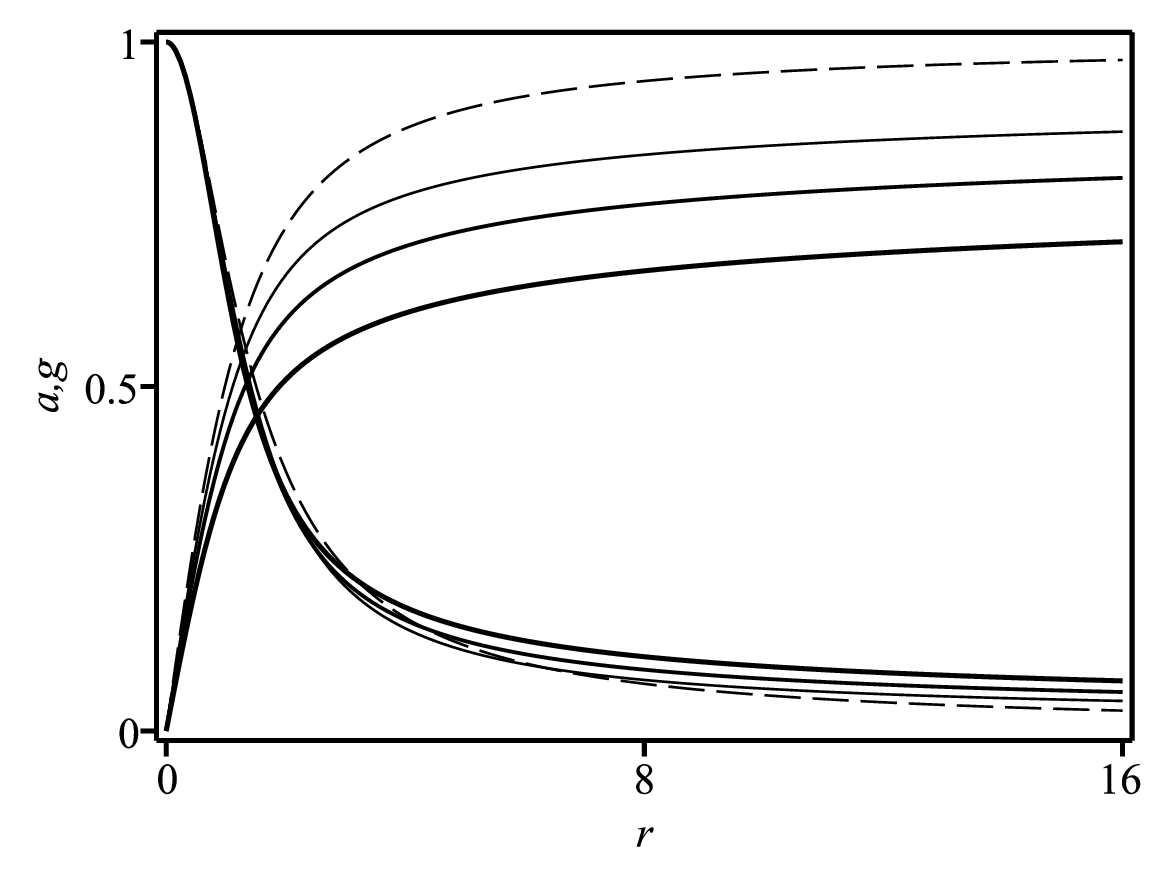}
\includegraphics[width=0.5\linewidth]{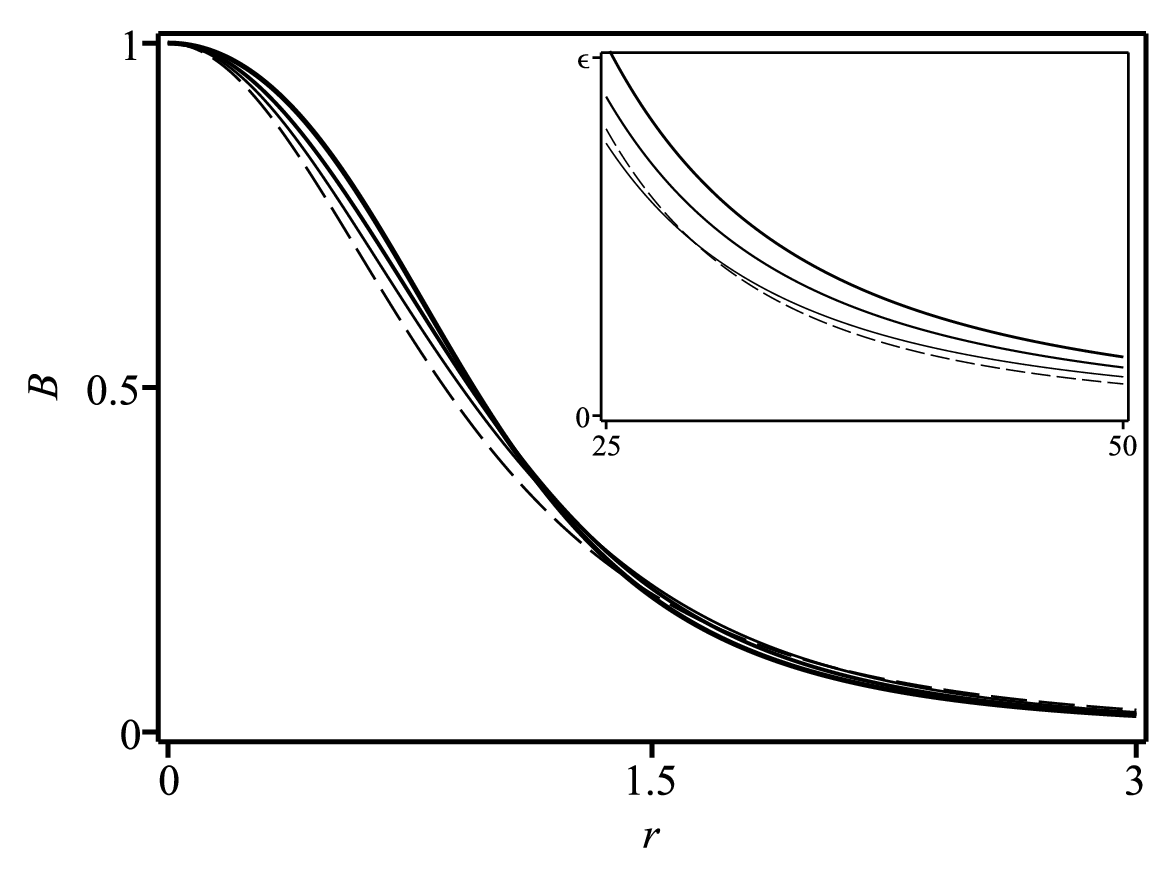}\includegraphics[width=0.5\linewidth]{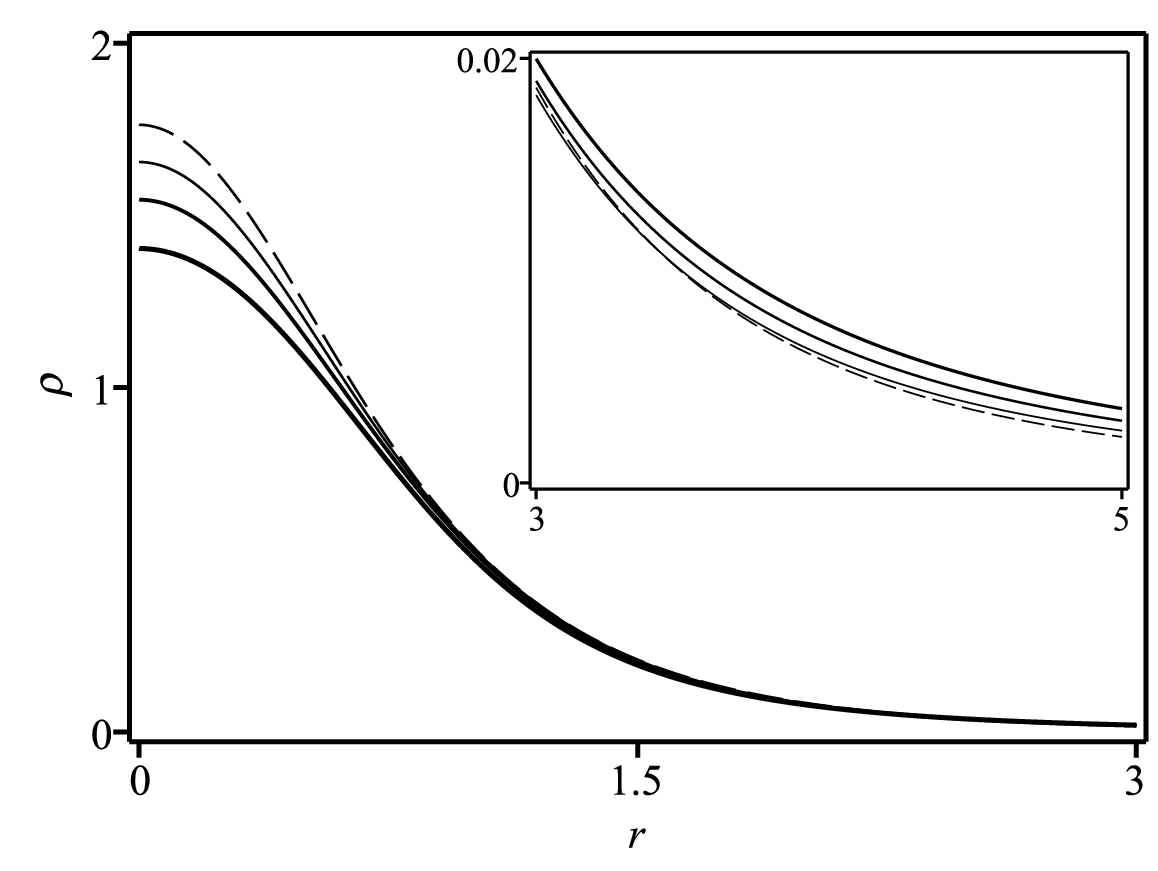}
\caption{The potential \eqref{potnum} (top left), the solution formed by the pair $a(r)$ (decreasing lines) and $g(r)$ (increasing lines) of the Eq.~\eqref{fonum} (top right), the magnetic field \eqref{bfield} (bottom left) and the energy density \eqref{rhoaz} (bottom right) for $b=0,1,2$ and $4$. The dashed lines stand for the case in which $b=0$. The insets show the tail for $r\in[25,50]$ in the magnetic field with $\epsilon=4\times10^{-5}$ and for $r\in[3,5]$ in the energy density. The thickness of the lines increases with $b$.}
\label{fig1}
\end{figure}

In Fig.~\ref{fig1}, we display the potential \eqref{potnum}, the solution of the first-order equations \eqref{fonum} formed by the pair $a(r)$ and $g(r)$, and the corresponding magnetic field \eqref{bfield} and energy density \eqref{rhoaz}. As expected, we see that the aforementioned physical quantities approach the long-range ones as $b$ gets near zero. The magnetic field and energy density are localized for all values of $b$, vanishing for $r\to\infty$. In the insets, we display their respective tail to show that, although the $B(r)$ and $\rho(r)$ seem to fall off equally at first glance, in fact the tails of the super long-range structure vanish slower than the power-law ones. Also, in the super long-range case, the aforementioned two quantities go slower and slower to zero as $b$ gets larger and larger. 

As we have commented above, the magnetic permeability in Eq.~\eqref{munum1} allows for the presence of super long-range vortices. It is worth remarking that we can generalize it to
\be
\mu(\vphia) =f(\vphia) \left|1-\vphia^2\right|^{2s}\sech^2\left(\frac{b\sqrt{s(s+1)/2}\,\vphia^2}{|1-\vphia^2|^{1/s}}\right),
\ee
for $\vphia\neq1$, and $\mu(1)=0$. In this situation, $s$ is a real parameter obeying $s>0$ and $f(\vphia)$ is an arbitrary limited non-negative function without zeroes in the interval $0<\vphia\leq1$. We have found by expanding the above expression around $\vphia=1$ that the correspondence with \eqref{musuper} is given by $\alpha=2\zeta$, $\beta=b$ and $\gamma=s$, where we have defined $\zeta=\sqrt{f(1)}$. So, the solution engender super long-range tails as in Eq.~\eqref{asysuper}, which reads
\be
g(r)\approx 1 -\frac{\vartheta}{\ln^{s}(2\zeta br)}\quad\text{and}\quad a(r)\approx \frac{s\vartheta}{\ln^{s+1}(2\zeta br)},
\ee
in which $\vartheta=(2^{-\frac{s+2}{s}}b^2s(s+1))^{s/2}$. Therefore, the presence of the parameter $s$ introduces an exponent in the logarithm that controls the strength of the falloff. As $s$ gets larger, the tails vanish faster, albeit still logarithmically.

\section{Second model}\label{secsecond}
We now introduce another class of models with a non-negative parameter $s$ which modifies the exponent of the logarithmic falloff and allows for the presence of analytical solutions. It is given by
\be\label{muana}
\begin{aligned}
    \mu(\vphia) &= \frac{\left|1-\vphia^{2s}\right|^{\frac{s+1}{s}}}{\vphia^2\,|1-\vphia^2|}\\
    &\times\left(1+2s\vphia^{2s} -\frac{2b\vphia C_b(\vphia)}{\left|1-\vphia^{2s}\right|^{\frac{1}{2s}}}\right)S_b^2(\vphia)
\end{aligned}
\ee
for $\vphia\neq1$, and $\mu(1)=0$, where
\bes\label{cbsb}
\bal
C_b(\vphia)&=\csch\left(\frac{2b\vphia}{\left|1-\vphia^{2s}\right|^{\frac{1}{2s}}}\right),\\
S_b(\vphia)&=\sech\left(\frac{b\vphia}{\left|1-\vphia^{2s}\right|^{\frac{1}{2s}}}\right).
\eal
\ees
By expanding the magnetic permeability \eqref{muana} around $\vphia=1$, one sees that it attains the very same form in Eq.~\eqref{musuper}, with
\be\label{correspondencia1}
\alpha=2\sqrt{2s+1}\,s^{\frac{s+1}{2s}},\quad \beta=\frac{b}{\sqrt{2s+1}s^{\frac{s+1}{2s}}},\quad  \gamma=2s.
\ee
Therefore, we expect it to support super long-range vortex configurations. By using Eq.~\eqref{pot}, we get the potential 
\be\label{potana}
\begin{aligned}
V(\vphia) &= \frac{1}{2\vphia^2}\left|1-\vphia^2\right|\left|1-\vphia^{2s}\right|^{\frac{s+1}{s}}\\
    &\times\left(1+2s\vphia^{2s} -\frac{2b\,\vphia\,C_b(\vphia)}{\left|1-\vphia^{2s}\right|^{\frac{1}{2s}}}\right)S_b^2(\vphia)
\end{aligned}
\ee
for $\vphia\neq1$ and $V(1)=0$. The above potential is non negative. The point $\vphia=0$ must be dealt with care, because it may lead to divergences in the derivatives of the potential. To avoid them, we take $s=1$ and $s\geq2$. For $s=1$, we get $V(0)=1+b^2/3$ and, for $s>1$, $V(0)=b^2/3$. By taking the limit $b\to0$, one recovers the long-range models previously investigated in Ref.~\cite{longrange3}.

For general $b$, the first-order equations \eqref{foaz} combined with the magnetic permeability \eqref{muana} read
\bes\bal
g^\prime &= \frac{ag}{r}, \\
-\frac{a^\prime}{r} &= \frac{\left|1-g^{2s}\right|^{\frac{s+1}{s}}}{g^2}\left(1+2sg^{2s} -\frac{2b\,g\,C_b(g)}{\left|1-g^{2s}\right|^{\frac{1}{2s}}}\right)S_b^2(g).
\eal
\ees
Considering the boundary conditions \eqref{bound} with $n=1$, they are solved by
\bes\label{solana}
\bal
g(r) &= \frac{\arcsinh(br)}{\left(b^{2s}+\arcsinh^{2s}(br)\right)^{1/(2s)}},\\
a(r) &= \frac{b^{2s+1}r}{\arcsinh(br)\big(b^{2s}+\arcsinh^{2s}(br)\big)\sqrt{1+b^2r^2}}.
\eal
\ees
The above pair of functions has logarithmic asymptotic behavior, in the form
\be\label{asyanas}
g(r) \approx 1-\frac{b^{2s}}{2s\ln^{2s}(2br)}\quad\text{and}\quad a(r) \approx \frac{b^{2s}}{\ln^{2s+1}(2br)},
\ee
as expected from the relation of the parameters $\alpha$, $\beta$ and $\gamma$ in \eqref{musuper} with $s$ and $b$ as in Eq.~\eqref{correspondencia1}. Notice that the correspondence between the parameters also relates the above expressions with \eqref{asysuper}. Therefore, the solution \eqref{solana} engenders super long-range tails.

In the limit $b\to0$, Eq.~\eqref{solana} reduces to the long range solution
\be
g(r) = \frac{r}{\left(1+r^{2s}\right)^{1/(2s)}} \quad\text{and}\quad a(r) = \frac{1}{1+r^{2s}},
\ee
which has power-law asymptotic behavior, $g(r) \approx 1 -1/(2sr^{2s})$ and $a(r) \approx 1/r^{2s}$.

By using Eq.~\eqref{bfield}, one can show that the magnetic field associated to the super long-range solution \eqref{solana} is
\be\label{bana}
\begin{aligned}
B(r) &= \frac{b^{2(s+1)}\left(b^{2s}+(2s+1)\,\arcsinh^2(br)\right)}{\arcsinh^2(br)\big(b^{2s}+\arcsinh^{2s}(br)\big)^2\big(1+b^2r^2\big)}\\
    &-\frac{b^{2s+1}}{r\,\arcsinh(br)\big(b^{2s}+\arcsinh^{2s}(br)\big)\big(1+b^2r^2\big)^{\frac32}},
\end{aligned}
\ee
which falls off asymptotically with a combination of power-law and logarithmic functions, as $B(r)\approx b^{2s}(2s+1)/(r^2\ln^{2(s+1)}(2br))$. The energy density \eqref{rhoaz} is
\be\label{rhoana}
\begin{aligned}
\rho(r) &= \frac{b^{2(s+1)}\left(b^{2s}+(2s+1)\,\arcsinh^{2s}(br)\right)}{\arcsinh^2(br)\big(b^2+\arcsinh^{2s}(br)\big)^2\big(1+b^2r^2\big)}\\
    &+ \frac{b^{2(s+1)}\left(b^{2s}-(2s+1)\,\arcsinh^{2s}(br)\right)}{\big(b^2+\arcsinh^{2s}(br)\big)^{\frac{2s+1}{s}}\big(1+b^2r^2\big)}\\
    &-\frac{b^{2s+1}\big(\arcsinh^2(br)-\big(b^2+\arcsinh^{2s}(br)\big)^{\frac1s}\big)}{r\,\arcsinh(br)\big(b^2+\arcsinh^{2s}(br)\big)^{\frac{s+1}{s}}\big(1+b^2r^2\big)^{\frac32}},
\end{aligned}
\ee
which behaves as $\rho(r)\approx b^{4s}(4s+1)/(sr^2\ln^{2(2s+1)}(2br))$ asymptotically. Notice that both the magnetic field and the energy density obeys Eqs.~\eqref{basysuper} and \eqref{rhoasysuper} for very large $r$ with the correspondence in~\eqref{correspondencia1}. Therefore, we see that the logarithmic tails of the solutions \eqref{solana} plays an important role in the physical properties of the vortex, as its magnetic field and energy density become modified, requiring a farther range to vanish. Similarly to what happens with the solutions, the limit $b\to0$ lead to the power-law profiles previously investigated in Ref.~\cite{longrange3}. By integrating \eqref{rhoana}, we have obtained energy $E=2\pi$, matching with the value expected from the first-order formalism for $n=1$.

\begin{figure}[t!]
\centering
\includegraphics[width=0.5\linewidth]{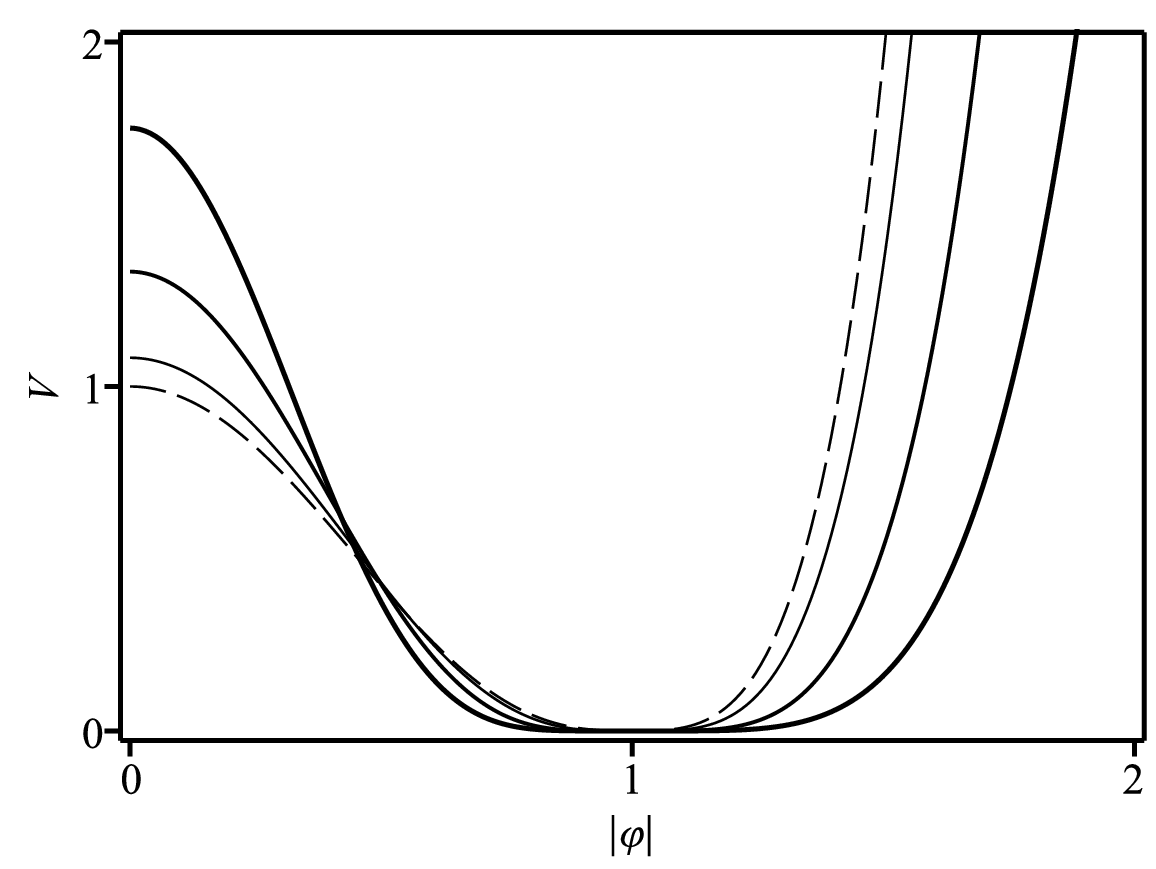}\includegraphics[width=0.5\linewidth]{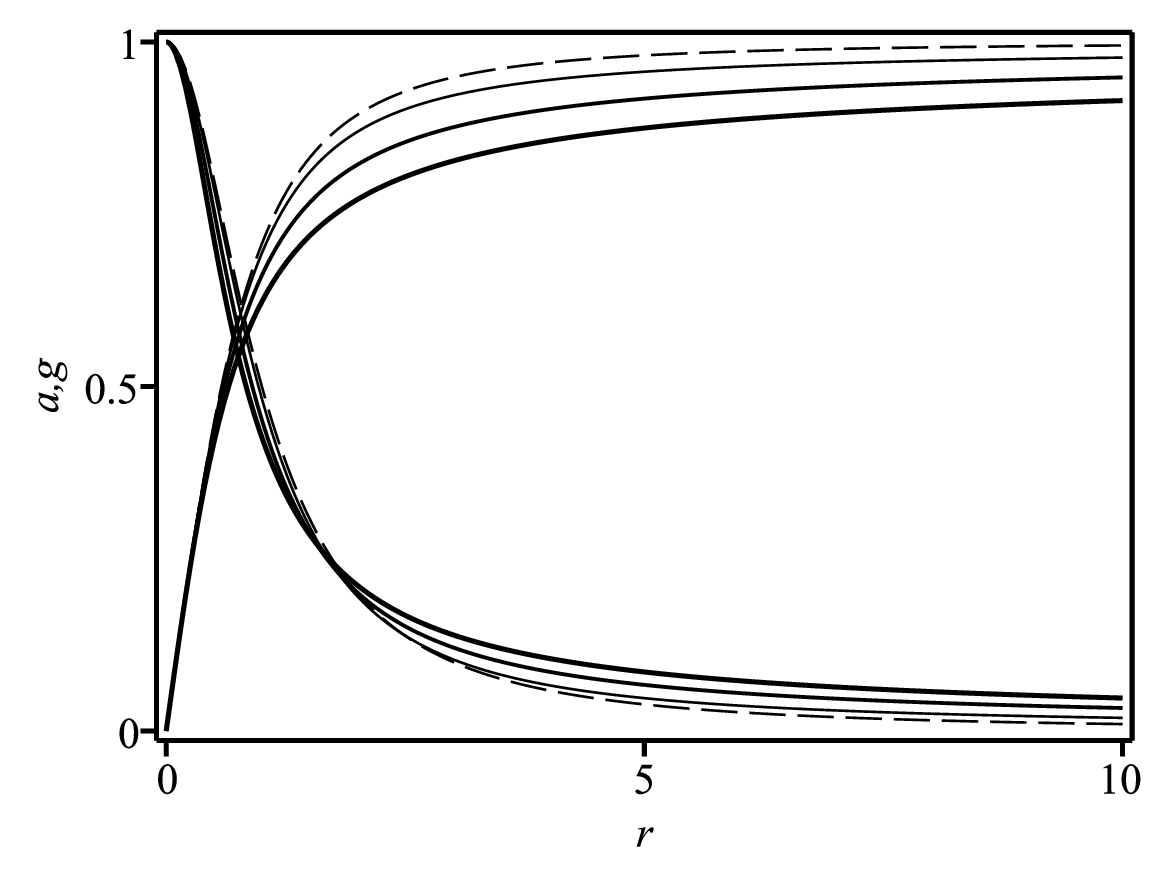}
\includegraphics[width=0.5\linewidth]{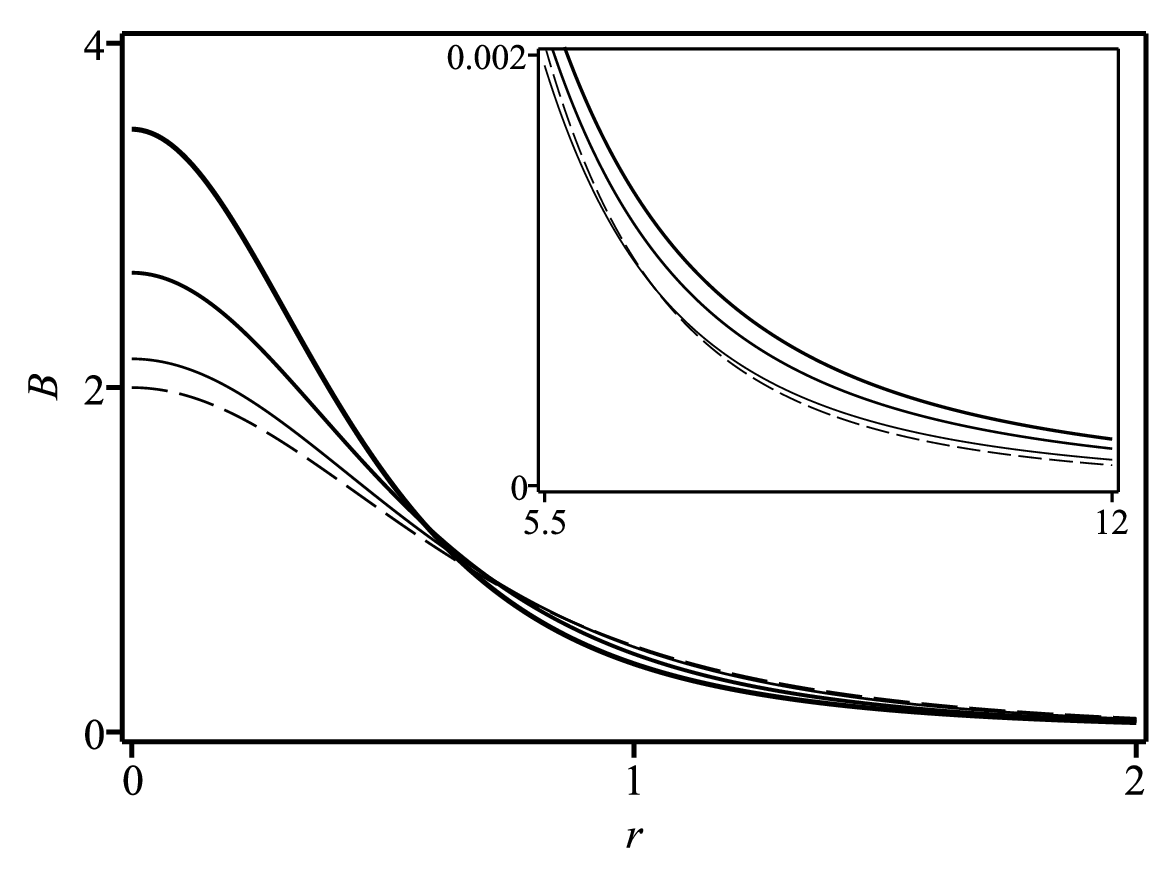}\includegraphics[width=0.5\linewidth]{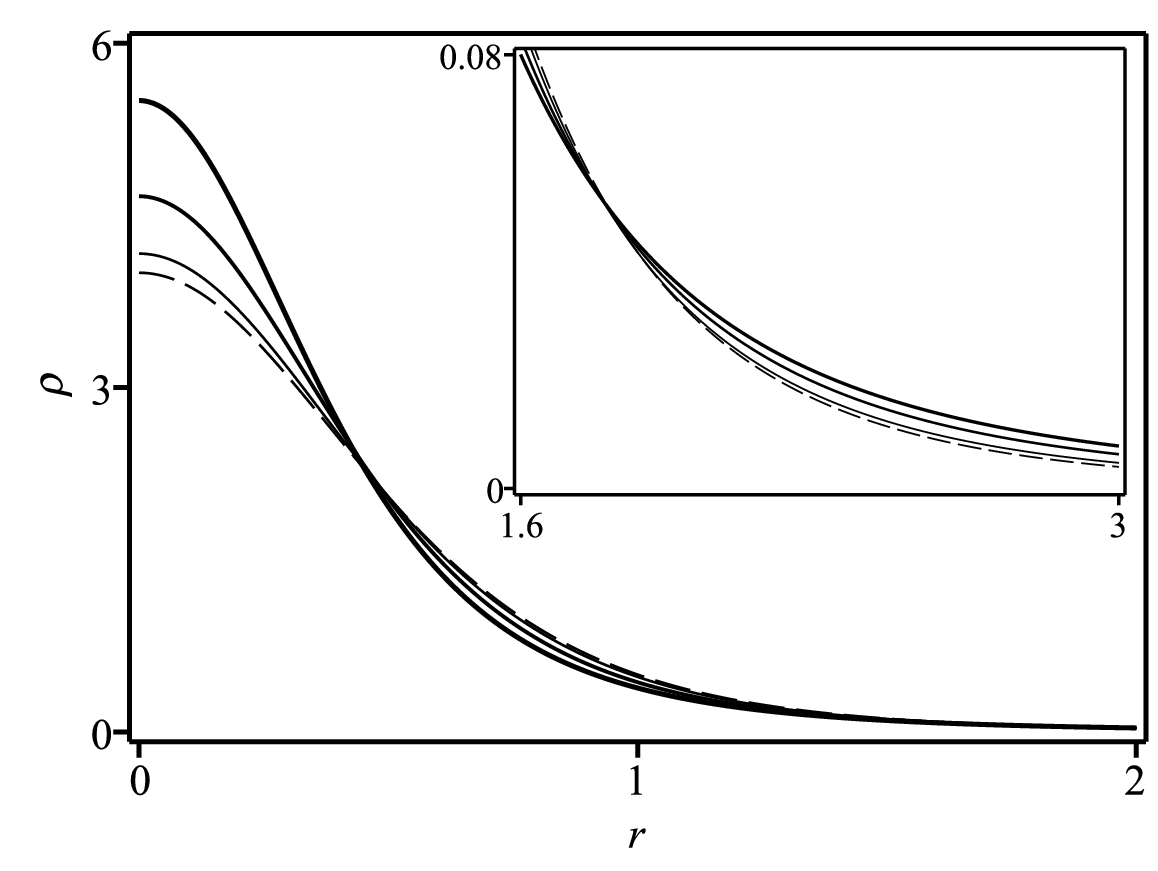}
\caption{The potential \eqref{potana} (top left), the solution formed by the pair $a(r)$ (decreasing lines) and $g(r)$ (increasing lines) in Eq.~\eqref{solana} (top right), the magnetic field \eqref{bana} (bottom left) and the energy density \eqref{rhoana} (bottom right) for the super long-range vortex, with $s=1$  and $b=1/2,1$ and $3/2$. The dashed lines stand for the long-range limit, $b\to0$. The insets show the tail for $r\in[5.5,12]$ in the magnetic field and for $r\in[1.6,3]$ in the energy density. The thickness of the lines increases with $b$.}
\label{fig2}
\end{figure}

To better visualize the super long-range vortex configuration described by the permeability \eqref{muana}, we display in Fig.~\ref{fig2} the potential \eqref{potana}, the solution \eqref{solana}, the magnetic field \eqref{bana} and the energy density \eqref{rhoana} for $s=1$ and some values of $b$, including the limit $b\to0$ whose solution is of power-law type. The minimum of the potential at $\vphia=1$ becomes wider and the solution falls off slower as $b$ gets larger. At first glance, the bottom plots seem to show that the tails are very similar for all the values of $b$. However, by looking at the insets, one sees that $B(r)$ and $\rho(r)$ for $b>0$ take a larger interval in $r$ to vanish than for $b=0$ and, as $b$ approaches zero, they become closer to the power-law limit, $b\to0$. Therefore, the parameter $b$ controls how close the logarithm and power-law decays are.

\begin{figure}[t!]
\centering
\includegraphics[width=0.5\linewidth]{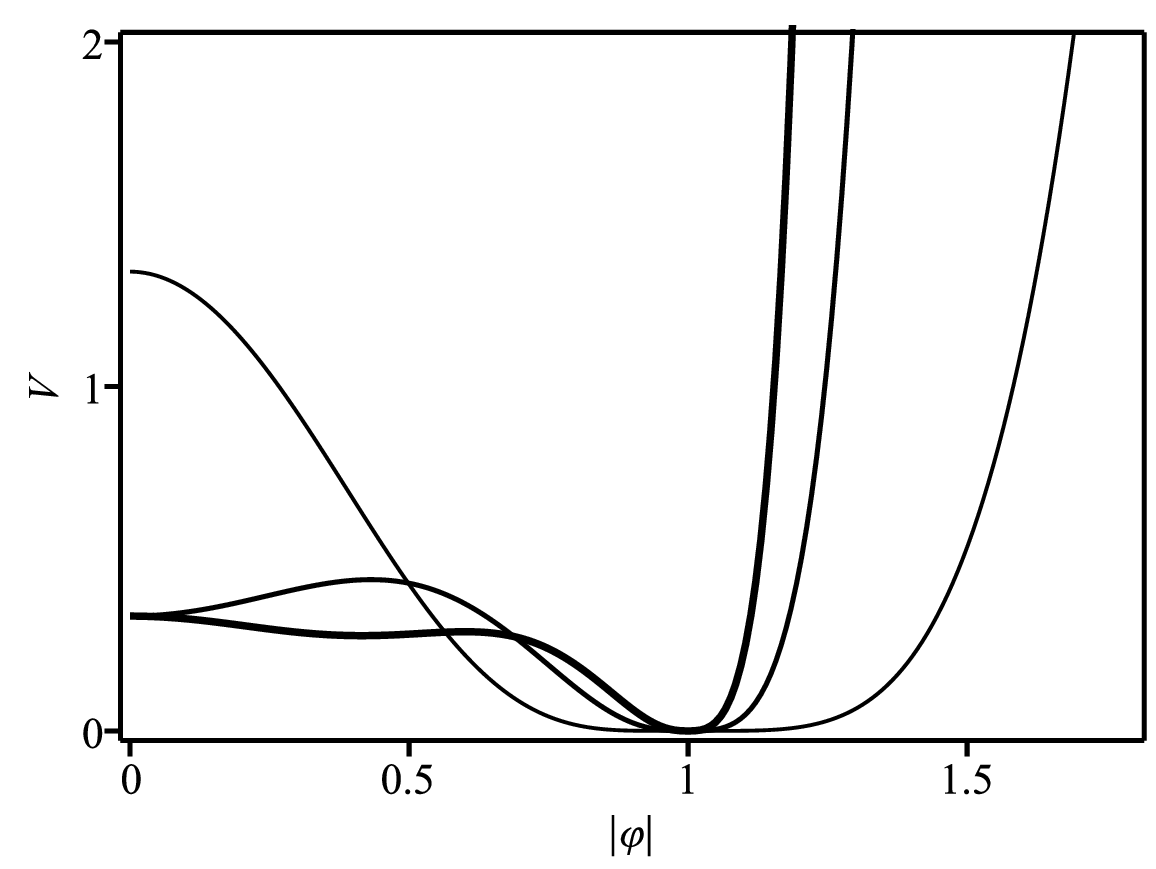}\includegraphics[width=0.5\linewidth]{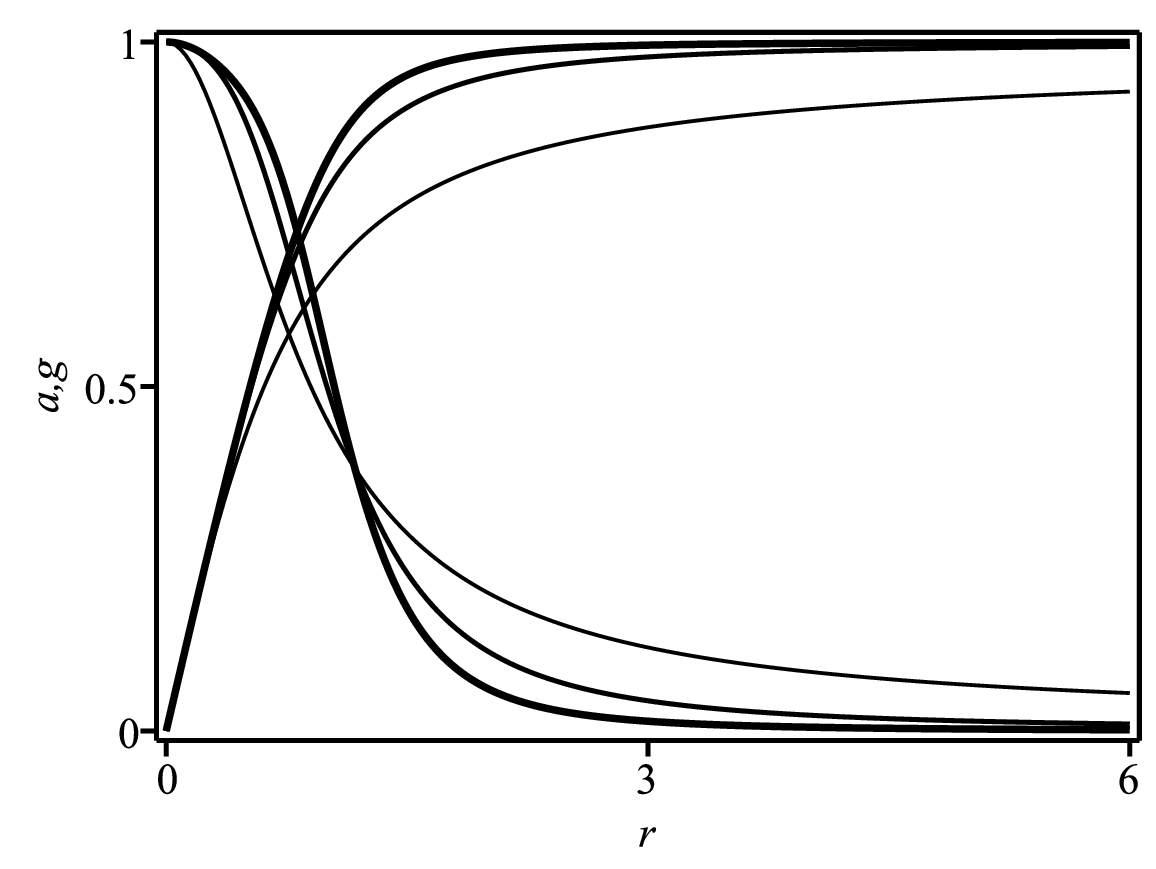}
\includegraphics[width=0.5\linewidth]{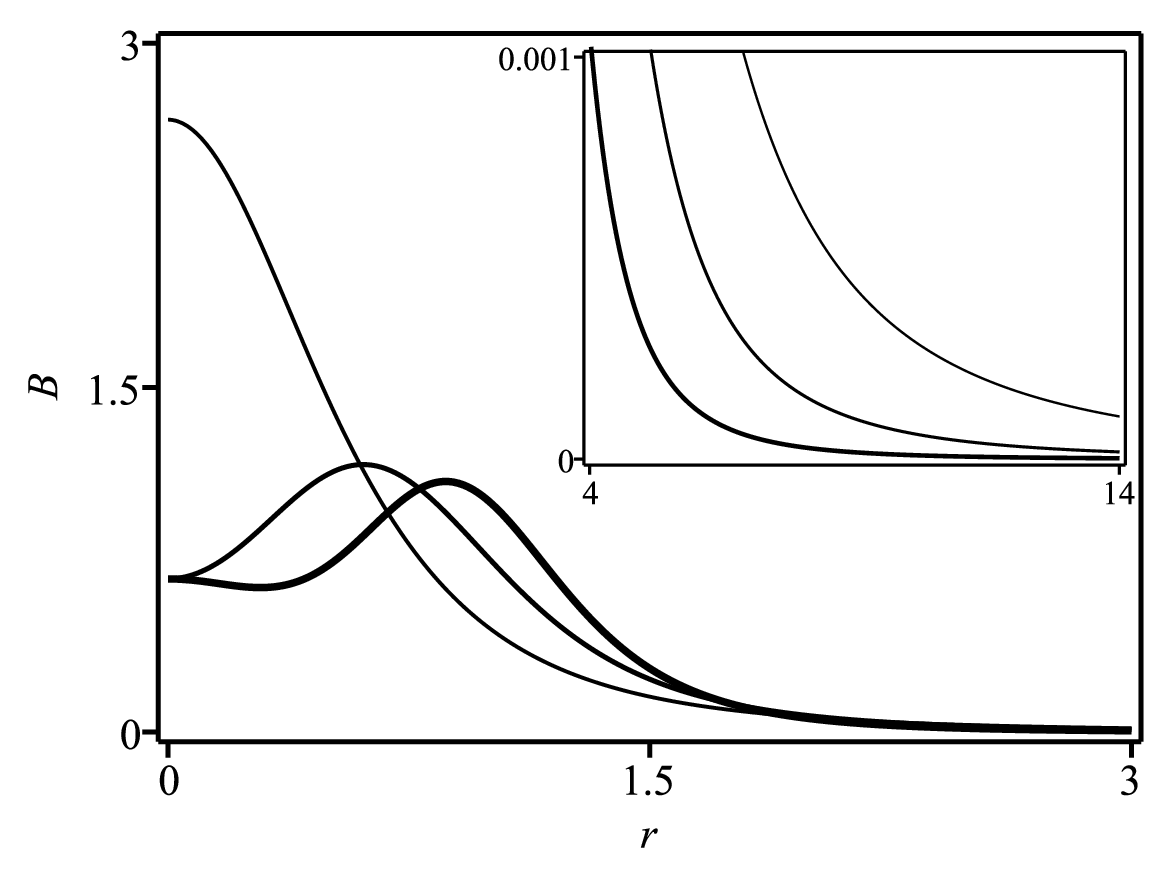}\includegraphics[width=0.5\linewidth]{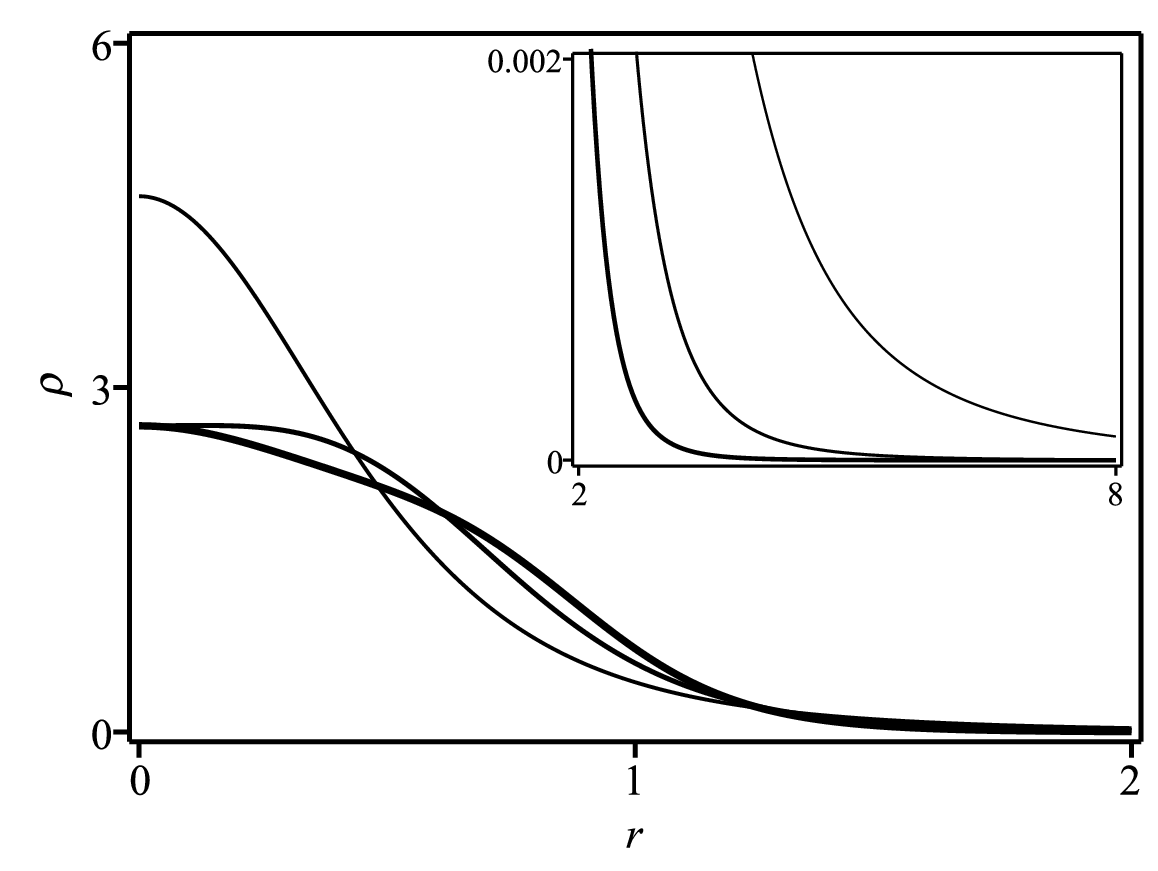}
\caption{The potential \eqref{potana} (top left), the solution formed by the pair $a(r)$ (decreasing lines) and $g(r)$ (increasing lines) in Eq.~\eqref{solana} (top right), the magnetic field \eqref{bana} (bottom left) and the energy density \eqref{rhoana} (bottom right) for $b=1$ and $s=1,2$ and $3$. The insets show the tail for $r\in[4,14]$ in the magnetic field and for $r\in[2,8]$ in the energy density. The thickness of the lines increases with $s$.}
\label{fig3}
\end{figure}

We remark that the model also has the parameter $s$, which appears in the exponent of the logarithmic functions that dictates the asymptotic behavior of the super long-range vortex solutions; see Eq.~\eqref{asyanas}. In Fig.~\ref{fig3}, we display the potential \eqref{potana}, the solution \eqref{solana}, the magnetic field \eqref{bana} and the energy density \eqref{rhoana} for $b=1$ and several values of $s$. We see that, the more $s$ increases, the faster the solution and its corresponding $B(r)$ and $\rho(r)$ vanish. Therefore, the parameter $s$ controls how fast the super long-range vortex falls off. It is worth emphasizing that, differently to what occurs when varying $b$, there is no limit for $s$ with $b>0$ that makes the logarithmic tails disappear.

\section{Outlook}\label{secoutlook}
In this work, we have investigated the presence of vortex solutions with logarithmic decay, which we call \emph{super long-range vortices}, in Maxwell-Higgs models with generalized magnetic permeability. We have used the Bogomol'nyi procedure to obtain first-order equations and potential that lead to minimum energy.

By analyzing the behavior at the vacua ($\vphia=1$) of the model, we have found a class of magnetic permeabilities that allow for the presence of the aforementioned feature. In the magnetic field and energy density, the logarithm that appears in the solution mixes with power-law functions. Interestingly, the magnetic permeability which we obtained engenders parameters which support limits that connect the super long-range to the standard or long-range solution.

To illustrate our procedure, we have gotten inspiration from Ref.~\cite{superlongkink2} and considered the magnetic permeability \eqref{munum1}. In this case, even though the first-order equations do not support analytical solutions, we have found that asymptotic behavior of numerical solutions falls off logarithmically, exhibiting the super long-range profile. We then have discussed a generalization that includes a parameter $s$ which modify the exponent of the logarithmic tail. In this direction, we have considered another class of models which leads to analytical solutions engendering the super long-range behavior.

As perspectives, one may consider the inclusion of a function which modifies the dynamical term of the scalar field in the Lagrangian density, as
\be
\LL = -\frac{1}{4\mu(\vphia)}F_{\alpha\beta}F^{\alpha\beta} +M(\vphia)\ov{D_\alpha\vphi}D^\alpha\vphi - V(\vphia).
\ee
The presence of $M(\vphia)$ is of interest, as it may provide a path to obtain super long-range vortices in the Chern-Simons-Higgs model
\be\label{modelCS}
\LL = \frac14\varepsilon^{\alpha\beta\gamma}A_\gamma F_{\alpha\beta} +M(\vphia)\ov{D_\alpha\vphi}D^\alpha\vphi - V(\vphia).
\ee
Notice that one cannot modify the Chern-Simons term with $\mu(\vphia)$ to keep gauge invariance, so the function $M(\vphia)$ is the only way with self-interactions of the field to get solutions with behavior different from the one obtained in Ref.~\cite{CSko,CSjak1}. In this direction, on may consider the class of models studied in Refs.~\cite{CSgeral1,CSgeral2,CSgeral3}, to verify if it admits super long-range vortex configurations.

Other aspects of vortices, such as collisions, interactions and decay, may also be possibilities for future research, following the lines of Refs.~\cite{colisaonew1,colisaonew2,colisaonew3,colisaonew4,colisaonew5}. Since the super long-range vortex engenders interactions that extend farther than the standard ones, novel features may arise in the aforementioned topics. Moreover, one may also investigate the interaction of this new structure with impurities \cite{imp1,imp2,imp3,imp4,imp5}. There is also the possibility of enhance the symmetry of the model \eqref{modelM} to accommodate more fields via an $U(1)^N$ symmetry. This may be of interest to include dark sectors in the model; see Refs.~\cite{dark1,dark2,dark3,dark4,dark5,dark6}.

\acknowledgments{We acknowledge financial support from the Brazilian agencies Conselho Nacional de Desenvolvimento Cient\'ifico e Tecnol\'ogico (CNPq), grants Nos. 402830/2023-7 (MAM and RM), 306151/2022-7 (MAM) and 310994/2021-7 (RM), and Paraiba State Research Foundation (FAPESQ-PB) grant No. 2783/2023 (IA).}



\begin{thebibliography}{99}
\bibitem{NO}
H.B.~Nielsen and P.~Olesen,
Nucl. Phys. B \textbf{61}, 45 (1973).

\bibitem{bogo}
E.B.~Bogomolny,
Sov. J. Nucl. Phys. \textbf{24}, 449 (1976).

\bibitem{schaposnik}
H.J.~de Vega and F.A.~Schaposnik,
Phys. Rev. D \textbf{14}, 1100 (1976).

\bibitem{vortexM1}
M.A.~Lohe,
Phys. Rev. D \textbf{23}, 2335 (1981).

\bibitem{vortexM2}
M.A.~Lohe and J.~van der Hoek,
J. Math. Phys. \textbf{24}, 148 (1983).

\bibitem{mu1}
J.H.~Lee and S.~Nam,
Phys. Lett. B \textbf{261}, 437 (1991).

\bibitem{mu2}
D.~Bazeia,
Phys. Rev. D \textbf{46}, 1879 (1992).

\bibitem{mu3}
D.~Bazeia, E.~da Hora, C.~dos Santos and R.~Menezes,
Eur. Phys. J. C \textbf{71}, 1833 (2011)
[arXiv:1201.2974 [hep-th]].

\bibitem{mu4}
A.N.~Atmaja, H.S.~Ramadhan and E.~da Hora,
JHEP \textbf{2016}, 117 (2016)
[arXiv:1505.01241 [hep-th]].

\bibitem{longrange1}
R.~Casana, M.M.~Ferreira, Jr., E.~da Hora and C.~dos Santos,
Adv. High Energy Phys. \textbf{2014}, 210929 (2014)
[arXiv:1405.7920 [hep-th]].

\bibitem{mu5}
D.~Bazeia, R.~Casana, M.M.~Ferreira and E.~da Hora,
EPL \textbf{109}, 21001 (2015)
[arXiv:1502.05063 [hep-th]].

\bibitem{mu56}
D.~Bazeia, L.~Losano, M.A.~Marques, R.~Menezes and I.~Zafalan,
Eur. Phys. J. C \textbf{77}, 63 (2017)
[arXiv:1611.02110 [hep-th]].

\bibitem{longrange2}
D.~Bazeia, L.~Losano, M.A.~Marques and R.~Menezes,
Phys. Lett. B \textbf{778}, 22 (2018)
[arXiv:1801.01077 [hep-th]].

\bibitem{ring1}
D.~Bazeia, M.~A.~Marques and D.~Melnikov,
Phys. Lett. B \textbf{785}, 454 (2018)
[arXiv:1807.02007 [hep-th]].

\bibitem{longrange3}
I.~Andrade, D.~Bazeia, M.A.~Marques and R.~Menezes,
Phys. Rev. D \textbf{102}, 025017 (2020)
[arXiv:2007.03138 [hep-th]].

\bibitem{ring2}
F.C.E.~Lima and C.A.S.~Almeida,
EPL \textbf{131}, 31003 (2020)
[arXiv:2007.12280 [hep-th]].

\bibitem{mu6}
F.C.E.~Lima, A.Y.~Petrov and C.A.S.~Almeida,
Phys. Rev. D \textbf{103}, 096019 (2021)
[arXiv:2101.09513 [hep-th]].

\bibitem{mu7}
I.~Andrade, M.A.~Marques and R.~Menezes,
EPL \textbf{133}, 31002 (2021)
[arXiv:2306.12542 [hep-th]].

\bibitem{mu8}
F.C.E.~Lima and C.A.S.~Almeida,
EPL \textbf{138}, 44001 (2022)
[arXiv:2202.06173 [hep-th]].

\bibitem{vortexY1}
E.~Babichev,
Phys. Rev. D \textbf{77}, 065021 (2008)
[arXiv:0711.0376 [hep-th]].

\bibitem{vortexY2}
D.~Bazeia, E.~da Hora, R.~Menezes, H.P.~de Oliveira and C.~dos Santos,
Phys. Rev. D \textbf{81}, 125016 (2010)
[arXiv:1004.3710 [hep-th]].

\bibitem{vortexY3}
D.~Bazeia, E.~da Hora and D.~Rubiera-Garcia,
Phys. Rev. D \textbf{84}, 125005 (2011)
[arXiv:1103.4940 [hep-th]].

\bibitem{vortexY4}
R.~Casana, E.~da Hora, D.~Rubiera-Garcia and C.~dos Santos,
Eur. Phys. J. C \textbf{75}, 380 (2015)
[arXiv:1507.08793 [hep-th]].

\bibitem{vortexY5}
R.~Casana, A.~Cavalcante and E.~da Hora,
JHEP \textbf{2016}, 051 (2016)
[arXiv:1509.04654 [hep-th]].

\bibitem{vortexY6}
A.N.~Atmaja,
Phys. Lett. B \textbf{768}, 351 (2017)
[arXiv:1511.01620 [hep-th]].

\bibitem{vortexY7}
D.~Bazeia, L.~Losano, M.A.~Marques, R.~Menezes and I.~Zafalan,
Nucl. Phys. B \textbf{934}, 212 (2018)
[arXiv:1708.07754 [hep-th]].

\bibitem{vortexY8}
F.C.E.~Lima, A.Y.~Petrov and C.A.S.~Almeida,
Phys. Rev. D \textbf{105}, 056005 (2022)
[arXiv:2109.15303 [hep-th]].

\bibitem{derrick} G.H. Derrick, 
J. Math. Phys. \textbf{5} 1252 (1964).

\bibitem{longkink1}
M.A.~Lohe,
Phys. Rev. D \textbf{20}, 3120 (1979).

\bibitem{longkink2}
D.K.~Campbell, M.~Peyrard and P.~Sodano,
Physica D \textbf{19}, 165 (1986).

\bibitem{longkink3}
M.~Mohammadi, N.~Riazi and A.~Azizi,
Prog. Theor. Phys. \textbf{128}, 615 (2012).

\bibitem{longkink4}
A.R.~Gomes, R.~Menezes and J.C.R.E.~Oliveira,
Phys. Rev. D \textbf{86}, 025008 (2012)
[arXiv:1208.4747 [hep-th]].

\bibitem{longkink5}
D.~Bazeia, R.~Menezes and D.C.~Moreira,
J. Phys. Comm. \textbf{2}, 055019 (2018)
[arXiv:1805.09369 [hep-th]].

\bibitem{longkink6}
I.~Andrade, M.A.~Marques and R.~Menezes,
Annals Phys. \textbf{473}, 169915 (2025)
[arXiv:2409.01961 [hep-th]].

\bibitem{colisao1}
N.S.~Manton,
J. Phys. A \textbf{52}, 065401 (2019)
[arXiv:1810.03557 [hep-th]].

\bibitem{colisao2}
I.C.~Christov, R.J.~Decker, A.~Demirkaya, V.A.~Gani, P.G.~Kevrekidis, A.~Khare and A.~Saxena,
Phys. Rev. Lett. \textbf{122}, 171601 (2019)
[arXiv:1811.07872 [hep-th]].

\bibitem{colisao3}
I.C.~Christov, R.J.~Decker, A.~Demirkaya, V.A.~Gani, P.G.~Kevrekidis and R.V.~Radomskiy,
Phys. Rev. D \textbf{99}, 016010 (2019)
[arXiv:1810.03590 [hep-th]].

\bibitem{colisao4}
I.C.~Christov, R.J.~Decker, A.~Demirkaya, V.A.~Gani, P.G.~Kevrekidis and A.~Saxena,
Commun. Nonlinear Sci. Numer. Simul. \textbf{97}, 105748 (2021)
[arXiv:2005.00154 [hep-th]].

\bibitem{colisao5}
J.G.F.~Campos and A.~Mohammadi,
Phys. Lett. B \textbf{818}, 136361 (2021)
[arXiv:2006.01956 [hep-th]].

\bibitem{colisao6}
D.~Bazeia, J.G.F.~Campos and A.~Mohammadi,
JHEP \textbf{2023}, 116 (2023)
[arXiv:2303.12482 [hep-th]].

\bibitem{colisao7}
J.G.F.~Campos and A.~Mohammadi,
JHEP \textbf{2024}, 056 (2024)
[arXiv:2309.05628 [hep-th]].

\bibitem{colisao8}
J.G.F.~Campos, A.~Mohammadi and T.~Romanczukiewicz,
JHEP \textbf{2025}, 166 (2025)
[arXiv:2411.12630 [hep-th]].

\bibitem{colisao9}
E.~Belendryasova, P.A.~Blinov, T.V.~Gani, A.A.~Malnev and V.A.~Gani,
Chaos, Solitons \& Fractals \textbf{194}, 116170 (2025).

\bibitem{superlongkink1}
V.A.~Gani and A.M.~Marjaneh,
EPL \textbf{148}, 24001 (2024)
[arXiv:2401.03569 [nlin.PS]].

\bibitem{superlongkink2}
I.~Andrade, M.A.~Marques and R.~Menezes,
Chaos, Solitons \& Fractals \textbf{192}, 116040 (2025)
[arXiv:2410.21241 [hep-th]].

\bibitem{CSko}
J.~Hong, Y.~Kim and P.Y.~Pac,
Phys. Rev. Lett. \textbf{64}, 2230 (1990).

\bibitem{CSjak1}
R.~Jackiw and E.J.~Weinberg,
Phys. Rev. Lett. \textbf{64}, 2234 (1990).

\bibitem{CSgeral1}
Y.~Yang,
Lett. Math. Phys. \textbf{23}, 179 (1991).

\bibitem{CSgeral2}
D.~Bazeia, E.~da Hora, C.~dos Santos and R.~Menezes,
Phys. Rev. D \textbf{81}, 125014 (2010)
[arXiv:1006.3955 [hep-th]].

\bibitem{CSgeral3}
C.~dos Santos and E.~da Hora,
Eur. Phys. J. C \textbf{70}, 1145-1151 (2010).

\bibitem{colisaonew1}
A.~Alonso-Izquierdo, W.G.~Fuertes, N.S.~Manton and J.~Mateos Guilarte,
JHEP \textbf{2024}, 020 (2024)

\bibitem{colisaonew2}
S.~Krusch, M.~Rees and T.~Winyard,
Phys. Rev. D \textbf{110}, 056050 (2024)
[arXiv:2406.04164 [math-ph]].

\bibitem{colisaonew3}
A.~Alonso-Izquierdo, J.~Mateos Guillarte, M.~Rees and A.~Wereszczynski,
Phys. Rev. D \textbf{110}, 065004 (2024)
[arXiv:2406.05725 [hep-th]].

\bibitem{colisaonew4}
A.~Alonso-Izquierdo, J.J.~Blanco-Pillado, D.~Migu\'elez-Caballero, S.~Navarro-Obreg\'on and J.~Queiruga,
Phys. Rev. D \textbf{110}, 065009 (2024)
[arXiv:2405.06030 [hep-th]].

\bibitem{colisaonew5}
A.~Alonso Izquierdo, N.S.~Manton, J.~Mateos Guilarte and A.~Wereszczynski,
Phys. Rev. D \textbf{110}, 085006 (2024)
[arXiv:2405.20249 [hep-th]].

\bibitem{imp1}
D.~Tong and K.~Wong,
JHEP \textbf{2014}, 090 (2014)
[arXiv:1309.2644 [hep-th]].
\bibitem{imp2}
A.~Cockburn, S.~Krusch and A.A.~Muhamed,
J. Math. Phys. \textbf{58}, 063509 (2017)
[arXiv:1512.01054 [hep-th]].
\bibitem{imp3}
J.~Ashcroft and S.~Krusch,
Phys. Rev. D \textbf{101}, 025004 (2020)
[arXiv:1808.07441 [hep-th]].
\bibitem{imp4}
D.~Bazeia, M.A.~Liao and M.A.~Marques,
Phys. Lett. B \textbf{825}, 136862 (2022)
[arXiv:2110.01956 [hep-th]].
\bibitem{imp5}
D.~Bazeia, J.G.F.~Campos and A.~Mohammadi,
JHEP \textbf{2024}, 108 (2024)
[arXiv:2404.11694 [hep-th]].

\bibitem{dark1}
P.~Arias and F.A.~Schaposnik,
JHEP \textbf{2014}, 011 (2014)
[arXiv:1407.2634 [hep-th]].

\bibitem{dark2}
P.~Arias, E.~Ireson, C.~N\'u\~nez and F.~Schaposnik,
JHEP \textbf{2015}, 156 (2015)
[arXiv:1410.7701 [hep-th]].

\bibitem{dark3}
P.~Arias, E.~Ireson, F.~A.~Schaposnik and G.~Tallarita,
Phys. Lett. B \textbf{749}, 368 (2015)
[arXiv:1505.06705 [hep-th]].

\bibitem{dark4}
D.~Bazeia, L.~Losano, M.A.~Marques and R.~Menezes,
Adv. High Energy Phys. \textbf{2019}, 3187289 (2019)
[arXiv:1805.07369 [hep-th]].

\bibitem{dark5}
D.~Bazeia, M.A.~Liao, M.A.~Marques and R.~Menezes,
Phys. Rev. Research. \textbf{1}, 033053 (2019)
[arXiv:1908.07871 [hep-th]].

\bibitem{dark6}
P.~Arias, A.~Arza, F.A.~Schaposnik, D.~Vargas-Arancibia and M.~Venegas,
Int. J. Mod. Phys. A \textbf{37}, 2250087 (2022)
[arXiv:2202.04138 [hep-th]].

\end{thebibliography}

\end{document}